\documentclass[conference]{IEEEtran}
\IEEEoverridecommandlockouts

\usepackage{cite}

\ifCLASSINFOpdf
   \usepackage[pdftex]{graphicx}
\else
\fi

\usepackage[cmex10]{amsmath}
\usepackage{array}
\usepackage{amsthm}
\usepackage{amsmath}
\usepackage{amsfonts}

\newtheorem{theorem}{Theorem}
\newtheorem{corollary}{Corollary}

\newtheorem{proposition}{Proposition}

\begin{document}

%
\title{Distributed Remote Vector Gaussian Source Coding with Covariance Distortion Constraints}


%
\author{\IEEEauthorblockN{
Adel Zahedi\IEEEauthorrefmark{1},
Jan \O stergaard\IEEEauthorrefmark{1},
S\o ren Holdt Jensen\IEEEauthorrefmark{1}, 
Patrick Naylor\IEEEauthorrefmark{2} and
S\o ren Bech\IEEEauthorrefmark{1}\IEEEauthorrefmark{3}}

\thanks{The research leading to these results has received funding from the European Union's Seventh Framework Programme (FP7/2007-2013) under grant agreement n$^\circ$ ITN-GA-2012-316969.}

\IEEEauthorblockA{\IEEEauthorrefmark{1}Department of Electronic Systems\\
Aalborg University, 9220 Aalborg, Denmark \\
Email: \{adz, jo, shj, sbe\}@es.aau.dk}
\IEEEauthorblockA{\IEEEauthorrefmark{2}Electrical and Electronic Engineering Department\\
London Imperial College, London SW7 2AZ, UK\\
Email: p.naylor@imperial.ac.uk}
\IEEEauthorblockA{\IEEEauthorrefmark{3}Bang \& Olufsen\\
7600 Struer, Denmark}
}

\maketitle

\begin{abstract}
In this paper, we consider a distributed remote source coding problem, where a sequence of observations of source vectors is available at the encoder. The problem is to specify the optimal rate for encoding the observations subject to a covariance matrix distortion constraint and in the presence of side information at the decoder. For this problem, we derive lower and upper bounds on the rate-distortion function (RDF) for the Gaussian case, which in general do not coincide. We then provide some cases, where the RDF can be derived exactly. We also show that previous results on specific instances of this problem can be generalized using our results. We finally show that if the distortion measure is the mean squared error, or if it is replaced by a certain mutual information constraint, the optimal rate can be derived from our main result.
\end{abstract}


\section{Introduction}

\subsection{Notation and Problem Statement}
\begin{figure}[!t]
\centering
\includegraphics[width=2.25in]{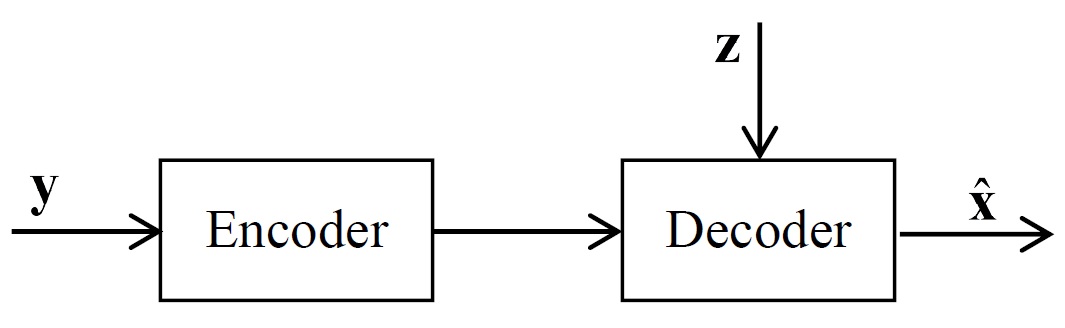}
\caption{Block diagram of the remote source coding problem}
\label{fig}
\end{figure}

We consider a stationary Gaussian source which generates independent vectors ${\bf{x}} \in {\mathbb{R}}^{n_x}$. A sequence of Gaussian vectors ${\bf{y}} \in {\mathbb{R}}^{n_y}$ which are measurements of the source is available at the encoder. Furthermore, a sequence of Gaussian vectors ${\bf{z}} \in {\mathbb{R}}^{n_z}$ is available at the decoder as side information. The problem is to specify the minimum rate for encoding $\bf{y}$ into a variable $\bf{u}$, so that the best estimation of the source from $\bf{u}$ and $\bf{z}$ at the decoder, denoted by $\bf{{\hat x}}$, satisfy a distortion constraint defined in form of a covariance matrix. This set-up is illustrated in Fig. \ref{fig}.

We denote conditional and nonconditional covariance and cross-covariance matrices by symbol $\bf{\Sigma}$ followed by an appropriate subscript. We assume that all covariance matrices are of full rank. Matrices and vectors are denoted by boldface uppercase and lowercase letters, respectively. A diagonal matrix having the elements $\lambda_1, ..., \lambda_n$ on its main diagonal is denoted by $\text{diag}\{\lambda_i, i=1,...,n\}$. Markov chains are denoted by two-headed arrows; e.g. ${\bf{y}} \leftrightarrow {\bf{x}} \leftrightarrow {\bf{z}}$, and the trace operation is denoted by $\text{tr}(\cdot)$. We use ${\bf{A} \succeq {\bf{B}}}$ to show that ${\bf{A} - {\bf{B}}}$ is positive semidefinite. Finally, we make use of the following notations:

\begin{eqnarray}
&&{(x)^ + } \buildrel \Delta \over = \max \left( {x,1} \right),\\
&&{(x)^ - } \buildrel \Delta \over = \min \left( {x,1} \right).
\end{eqnarray}

Using our notational convention, the problem described above can be formulated as specifying a rate-distortion function (RDF) $R(\bf{D})$, defined as:

\begin{eqnarray}
\label{mut-info}
&&R(\bf{D})=\min_{\bf{u} \in \mathcal{U}} \it{I}\left(\bf{y};\bf{u}|\bf{z}\right)\\
\label{constraint}
&&\text{subject to   } E\left[ {\left( {{\bf{x - \hat x}}} \right){{\left( {{\bf{x - \hat x}}} \right)}^T}} \right] {\preceq}\: {\bf{D}},
\end{eqnarray}

\noindent
where $\bf{D}$ is a symmetric positive-definite matrix specifying the target distortion, $\bf{\hat{x}}$ is defined as:

\begin{equation}
\label{xhat}
{\bf{\hat{x}}} = \it{E}\left[ \bf{x} | \bf{u},\bf{z} \right],
\end{equation}

\noindent
and $\mathcal{U}$ is the set of random variables $\bf{u}$ satisfying ${\bf{u}} \leftrightarrow {\bf{y}} \leftrightarrow {\left(\bf{x},\bf{z}\right)}$.

Following \cite{Tian}, and for simplicity of derivations we write the Gaussian vectors $\bf{x}$ and $\bf{y}$ in terms of linear estimation from other Gaussian vectors and estimation errors as follows:

\begin{eqnarray}
\label{estxzu}
&&\bf{x}=\bf{Az+Bu}+\bf{n}_1,\\
\label{estxyz}
&&\bf{x}=\bf{Cy+Gz}+\bf{n}_2,\\
\label{estyz}
&&\bf{y}=\bf{\bf{\Gamma }z}+\bf{n}_3,
\end{eqnarray}

\noindent
where $\bf{A}$, $\bf{B}$, $\bf{C}$, $\bf{G}$ and $\bf{\Gamma }$ are the coefficients of linear estimation, depending only on the covariance and cross-covariance matrices of $\bf{x, y, z}$ and $\bf{u}$, and ${\bf{n}}_i$, $i=1,2,3$ are estimation errors with covariance matrices ${\bf{\Sigma}}_{{\bf{n}}_1}$, ${\bf{\Sigma}}_{{\bf{n}}_2}={{\bf{\Sigma }}_{{\bf{x}}|{{\bf{yz}}}}}$, and ${\bf{\Sigma}}_{{\bf{n}}_3}={{\bf{\Sigma }}_{{\bf{y}}|{{\bf{z}}}}}$, respectively. (See the Appendix in \cite{myDCC} for more details.)

\subsection{Applications}
One possible application of the formulated problem is in wireless acoustic sensor networks which is a set of wireless microphones equipped with communication and signal processing units. The microphones are randomly distributed in an environment, sampling the sound field. The measurements made by each microphone should be delivered at a fusion center possibly via a sequence of transmissions through neighboring nodes. It is desirable to compute the local RDF at each node of such a network. This can be used for computing the network sum-rate,  by which one can formulate a sum-rate minimization problem with a distortion constraint (e.g. distortion at the final destination). As suggested in \cite{Jan}, this can be used for optimal routing in the network. When sending a message from a node to a neighboring node, the measurement of the sound at the latter node can be used as side information, thus fitting our distributed source coding framework. 

Another application is in relay networks as discussed in \cite{Tian}. In this case, $n_x$, $n_y$, and $n_z$ are the number of transmitter, relay, and receiver antennas.


\subsection{Related Work}
In the special case where $n_x = n_y = n_z = 1$, the RDF for the above-mentioned problem is given by \cite{Wornell}:

\begin{equation}
\label{scalar}
R(D) = \left\{
\begin{array}{l l}
\frac{1}{2}\log \left( {\frac{{{\Sigma _{x|z}} - {\Sigma _{x|yz}}}}{{D - {\Sigma _{x|yz}}}}} \right), & \quad \text{if } \Sigma _{x|yz} < D \le \Sigma _{x|z}\\
0, & \quad \text{if } D > \Sigma _{x|z}
\end{array} \right.
\end{equation}

\noindent
where $D$ is the scalar distortion constraint, and ${\Sigma _{x|z}}$ and ${\Sigma _{x|yz}}$ are conditional variances of scalar random variable $x$.

For the vector case, the authors in \cite{Tian} solved the problem for the mean-squared error distortion constraint; i.e. for problem (\ref{mut-info}) with the distortion constraint (\ref{constraint}) replaced by:

\begin{equation}
\label{mse}
{\rm{tr}}\left( {E\left[ {\left( {{\bf{x - \hat x}}} \right){{\left( {{\bf{x - \hat x}}} \right)}^T}} \right]} \right) \le {n_x}D.
\end{equation}

\noindent
This is similar to a case where in (\ref{constraint}), only the sum of the diagonal elements of the distortion and error matrices are of interest. Thus, in this particular case, the vector Gaussian problem can be treated as parallel scalar problems, leading to well-known water-filling interpretations. The RDF for this problem when $\text{tr} ({{\bf{\Sigma} _{x|yz}}}) \leq n_x D \leq \text{tr} ({{\bf{\Sigma} _{x|z}}})$ was shown in \cite{Tian} to be:

\begin{equation}
\label{trace}
R(D) = \frac{1}{2}\sum_{i=1}^{n_x} {\log {{\left( {\frac{{{\lambda _i}}}{\lambda }} \right)}^ + }} ,
\end{equation}

\noindent
where $\lambda _i; \: i = 1,2,...,n_x$ are the eigenvalues of ${\bf{C}}{{\bf{\Sigma }}_{{\bf{y|z}}}}{{\bf{C}}^{{T}}}$ with ${\bf{C}}$ defined in (\ref{estxyz}), and $\lambda$ satisfying the following constraint:

\begin{equation}
\label{water}
\sum\limits_{i = 1}^{{n_x}} {\min \left( {\lambda ,{\lambda _i}} \right)}  = {n_x}D - {\rm{tr}}\left( {{\bf{\Sigma} _{x|yz}}} \right).
\end{equation}

Related to our problem is also another problem considered in \cite{Tian}, where the constraint (\ref{constraint}) is replaced by a mutual information constraint defined as:

\begin{equation}
\label{mut-inf-const}
\it{I}\left(\bf{x};\bf{u}|\bf{z}\right) \geq R_I,
\end{equation}

\noindent
where $R_I$ is a given rate. The rate-rate function for this problem for $0 \leq R_I \leq \frac{1}{2}\log \left( {\frac{{\left| {{{\bf{\Sigma }}_{x|{{\bf{z}}}}}} \right|}}{{\left| { {{\bf{\Sigma }}_{x|{\bf{y}}{{\bf{z}}}}}} \right|}}} \right)$ is then given by:

\begin{equation}
\label{rate-rate}
{R}\left( {R_I} \right) = \frac{1}{2}\sum_{i=1}^{n_x} { \log \left( {\mu _i} {\left[ {\left( \frac{1 - {\mu _i}}{1 - \gamma} \right)}^{-} - \left( 1 - {\mu _i} \right) \right] }^ {-1} \right) },
\end{equation}

\noindent
where ${\mu _i}$, $i=1,...,n_x$ are the eigenvalues of ${{\bf{\Sigma}}_{{\bf{y|z}}}^{1/2}} {{\bf{C}}^T} {{\bf{\Sigma}}_{{\bf{x|z}}}^{-1}} {\bf{C}} {{\bf{\Sigma}}_{{\bf{y|z}}}^{1/2}}$, and $\gamma \in [0,1)$ satisfies the following:

\begin{eqnarray}
\label{mut-inf-water}
{-\frac{1}{2}} \sum_{i=1}^{n_x} {\log {\left( \frac{1 - {\mu _i}}{1 - \gamma} \right)}^{-} } = R_I.
\end{eqnarray}

In general, it is not straightforward to generalize the above results to the case of covariance matrix distortion constraints. Indeed, due to the matrix form of the distortion constraint, it does not appear as it is possible to reduce the problem to an equivalent problem of parallel scalar sources.

In \cite{myDCC}, the RDF for (\ref{constraint}) was recently found for the somewhat restrictive case where $n_x = n_y = n_z$, and ${\bf{C}}$ in (\ref{estxyz}) is invertible, and the distortion constraint satisfies ${{\bf{\Sigma }}_{{\bf{x}}|{{\bf{yz}}}}} \prec {\bf{D}} \preceq {{\bf{\Sigma }}_{{\bf{x}}|{{\bf{z}}}}}$. Under these assumptions, the RDF was shown to be:

\begin{equation}
\label{rd-dcc}
{R}\left( {{{\bf{D}}}} \right) = \frac{1}{2}\log \left( {\frac{{\left| {{{\bf{\Sigma }}_{x|{{\bf{z}}}}} - {{\bf{\Sigma }}_{x|{\bf{y}}{{\bf{z}}}}}} \right|}}{{\left| {{{\bf{D}}} - {{\bf{\Sigma }}_{x|{\bf{y}}{{\bf{z}}}}}} \right|}}} \right).
\end{equation}

Although under the above assumptions the problem is manageable to solve, it is a quite restricted case. In this paper, we consider the most general case with ${{\bf{\Sigma }}_{{\bf{x}}|{{\bf{yz}}}}} \prec {\bf{D}}$ and without the above assumptions, and establish a lower bound and an upper bound on the RDF, which in general do not coincide. Then we consider some special cases for which the two bounds coincide, giving the exact RDF. We will show that (\ref{scalar}) and (\ref{trace}) could be derived as special cases of our results. In addition, in the special case that $\bf{y}$ and $\bf{z}$ are noisy versions of $\bf{x}$ with additive white noise, (\ref{rate-rate}) could also be derived from our results. We will also generalize (\ref{rd-dcc}) to the case that no assumption is made on dimensions of vectors or invertibility of matrix $\bf{C}$.

The paper is organized as follows. Section II is dedicated to a brief presentation of some results from matrix algebra which will be used in our derivations. In Section III, we derive the lower and upper bounds on the RDF for the problem formulated above. In Section IV, we will establish the link between our results and (\ref{scalar}), (\ref{trace}), (\ref{rate-rate}), and (\ref{rd-dcc}). The paper is concluded in Section V.


\section{Simultaneous Diagonalization}
The following theorem is a weakened variant of Theorem 8.3.1 in \cite{matrix}, and will be the basis for some of the derivations in this work:

\begin{theorem}\label{th1}
For two symmetric positive definite $n \times n$ matrices ${\bf{\Sigma}}_1$ and ${\bf{\Sigma}}_2$, there is a nonsigular matrix $\bf{S}$ so that:

\begin{eqnarray}
\label{S1}
&&{\bf{S}} {\bf{\Sigma}}_1 {\bf{S}}^T = {\bf{I}}_n, \\
\label{S2}
&&{\bf{S}} {\bf{\Sigma}}_2 {\bf{S}}^T = \bf{\Gamma},
\end{eqnarray}

\noindent
where ${\bf{I}}_n$ is the identity $n \times n$ matrix, and $\bf{\Gamma}$ is a positive-definite $n \times n$ diagonal matrix.

\end{theorem}

Let us denote the eigenvalue decomposition of ${\bf{\Sigma}}_1$ by: 

\begin{equation}
\label{eig}
{\bf{\Sigma}}_1 = {\bf{U}}^T \bf{\Lambda U}.
\end{equation}

\noindent
We define the joint diagonalizer of ${\bf{\Sigma}}_1$ and ${\bf{\Sigma}}_2$ as:

\begin{equation}
\label{V}
{\bf{V}} = {\bf{\Lambda}}^{1/2} {\bf{S}}.
\end{equation}

\noindent
Using (\ref{V}) and (\ref{S1})--(\ref{S2}) we have:

\begin{eqnarray}
\label{V1}
&&{\bf{V}} {\bf{\Sigma}}_1 {\bf{V}}^T = \bf{\Lambda}, \\
\label{V2}
&&{\bf{V}} {\bf{\Sigma}}_2 {\bf{V}}^T = \bf{\Lambda}',
\end{eqnarray}

\noindent
where $\bf{\Lambda}'$ is defined as $\bf{\Lambda}' = \bf{\Lambda} \bf{\Gamma}$. Note that the diagonal elements in $\bf{\Lambda}'$ are not necessarily the eigenvalues of ${\bf{\Sigma}}_2$. However, if ${\bf{\Sigma}}_1$ and ${\bf{\Sigma}}_2$ commute, it is possible to find a joint eigenvalue decomposition for the two matrices, so that the matrix ${\bf{V}}$ in (\ref{V1}) and (\ref{V2}) is orthogonal, and $\bf{\Lambda}'$ consists of the eigenvalues of ${\bf{\Sigma}}_2$.

We will also make use of the following theorem (see \cite{matrix}, Theorem 8.4.9):

\begin{theorem}\label{th2}
Consider two positive semidefinite matrices ${\bf{Q}}_1$ and ${\bf{Q}}_2$, with eigenvalues $\lambda_1, ..., \lambda_n$ and $\mu_1,...,\mu_n$, respectively, which are sorted in order of magnitude. If ${\bf{Q}}_1 \succeq {\bf{Q}}_2$, then $\lambda_i \geq \mu_i$, for $i=1,...,n$.
\end{theorem}


\section{Main Results}
Let us define the matrices ${\bf{\Sigma}}_1$ and ${\bf{\Sigma}}_2$ as:

\begin{eqnarray}
\label{sig1}
&& {\bf{\Sigma}}_1 = {{{\bf{\Sigma }}_{\bf{x|z}}} - {{\bf{\Sigma }}_{\bf{x|yz}}}}, \\
\label{sig2}
&& {\bf{\Sigma}}_2 = {{\bf{D}} - {{\bf{\Sigma }}_{\bf{x|yz}}}},
\end{eqnarray}

\noindent
and denote their simultaneous diagonalization by ${\bf{\Lambda}}={\text{diag}}\{\lambda_i, i=1,...,n_x\}$ and ${\bf{\Lambda}}'={\text{diag}}\{\lambda'_i, i=1,...,n_x\}$, respectively. The eigenvalue decomposition of ${\bf{\Sigma}}_1$ is defined in (\ref{eig}). We also denote by $\lambda_{(i)}$ and $\lambda'_{(i)}$, $i=1,...,n_x$, the sorted-by-magnitude versions of $\lambda_i$ and $\lambda'_i$, $i=1,...,n_x$, respectively. The following theorem is the main result of this paper:

\begin{theorem}\label{th3}
The RDF formulated in (\ref{mut-info})--(\ref{constraint}) is bounded as follows:

\begin{equation}
\label{RD}
\frac{1}{2}{\sum_{i=1}^{n_x} {\log \left( {\frac{{{\lambda _{(i)}}}}{{{{\lambda '}_{(i)}}}}} \right)} ^ + } \leq R\left( {\bf{D}} \right) \leq  \frac{1}{2}{\sum_{i=1}^{n_x} {\log \left( {\frac{{{\lambda _i}}}{{{{\lambda}_i'}}}} \right)} ^ + }
\end{equation}

\end{theorem}

\begin{IEEEproof}
The proof follows from the results of the next two subsections. We first propose a scheme which achieves the upper bound. Then we prove that the RDF can be lower-bounded as in (\ref{RD}). 
\end{IEEEproof}


\subsection{Upper Bound}
\label{upper}
We will show that the upper bound in (\ref{RD}) is achievable by the following scheme:

\begin{equation}
\label{scheme}
{\bf{u}} = {\bf{UCy}} + {\bf{\nu}},
\end{equation}

\noindent
where the covariance matrix of the coding noise $\bf{\nu}$ is defined as:

\begin{equation}
\label{cov_noise}
{{\bf{\Sigma }}_{\bf{\nu}}} = {{\bf{UV}}^{-1}}{\rm{diag}}\left\{\! { \frac{{\lambda _i}\min \left({\lambda _i},{\lambda' _i}\right)}{{\lambda _i} \!-\! \min \left({\lambda _i},{\lambda' _i} \right)} } , i=1,...,n_x \! \right\}\!{{{\bf{V}}^{-T}}{{\bf{U}}^{T}}}.
\end{equation}

\begin{IEEEproof}
First notice that using (\ref{estxyz}) we can write:

\begin{equation}
\label{sch1}
{\bf{C}}{{\bf{\Sigma }}_{\bf{y|z}}}{{\bf{C}}^T} = {{{\bf{\Sigma }}_{\bf{x|z}}} - {{\bf{\Sigma }}_{\bf{x|yz}}}}.
\end{equation}

\noindent
Starting from ${I\left( {{\bf{y}};{\bf{u}}|{\bf{z}}} \right)}=h({\bf{u}}|{\bf{z}}) - h({\bf{u}}|{\bf{y}},{\bf{z}})$ and using (\ref{scheme}), (\ref{estyz}), and (\ref{sch1}), it is straightforward to show that:

\begin{equation}
\label{sch2}
I\left( {{\bf{y}};{\bf{u}}|{\bf{z}}} \right) = \frac{1}{2}\log \left( {\frac{{\left| {{\bf{\Lambda }} + {{\bf{\Sigma }}_{\bf{\nu}}}} \right|}}{{\left| {{{\bf{\Sigma }}_{\bf{\nu}}}} \right|}}} \right).
\end{equation}

\noindent
Noting that 

\begin{equation}
\label{lambda}
{{\bf{UV}}^{-1}}{\bf{\Lambda}}{{{\bf{V}}^{-T}}{{\bf{U}}^{T}}} = {\bf{\Lambda}},
\end{equation}

\noindent
and substituting (\ref{cov_noise}) in (\ref{sch2}) yields:

\begin{eqnarray}
I\left( {{\bf{y}};{\bf{u}}|{\bf{z}}} \right) &&= \frac{1}{2}\sum\limits_{i=1}^{n_x} {\log \left( {\frac{{{\lambda _i}}}{\min \left({\lambda _i},{\lambda' _i} \right)}} \right)} \nonumber \\
 &&= \frac{1}{2}\sum_{i=1}^{n_x} {\log {{\left( {\frac{{{\lambda _i}}}{{{\lambda '_i}}}} \right)}^ + }}.
\end{eqnarray}

Now we will show that using the coding scheme (\ref{scheme}) the reconstruction error at the decoder satisfies the distortion constraint (\ref{constraint}). First notice that form (\ref{estxzu}) it follows that ${{\bf{\Sigma }}_{xu|z}} = {\bf{B}}{{\bf{\Sigma }}_{u|z}}$, or:

\begin{equation}
\label{weird1}
{\bf{B}} = {{\bf{\Sigma }}_{xu|z}}{{\bf{\Sigma }}_{u|z}}^{ - 1}.
\end{equation}

\noindent
From (\ref{scheme}) and (\ref{sch1}) we have:

\begin{equation}
\label{weird2}
{{\bf{\Sigma }}_{u|z}} = {\bf{\Lambda }} + {{\bf{\Sigma }}_\nu}.
\end{equation}

\noindent
Also:

\begin{eqnarray}
\label{weird3}
{{\bf{\Sigma }}_{xu|z}} &&= {{\bf{\Sigma }}_{xy|z}}{{\bf{C}}^T}{{\bf{U}}^T}\\
\label{weird4}
 &&= {\bf{C}}{{\bf{\Sigma }}_{y|z}}{{\bf{C}}^T}{{\bf{U}}^T}\\
\label{weird5}
 &&= \left( {{{\bf{\Sigma }}_{x|z}} - {{\bf{\Sigma }}_{x|yz}}} \right){{\bf{U}}^T},
\end{eqnarray}

\noindent
where (\ref{weird3}), (\ref{weird4}) and (\ref{weird5}) follow from (\ref{scheme}), (\ref{estxyz}) and (\ref{sch1}), respectively. The covariance matrix of the reconstruction error can then be written as:

\begin{align}
\label{err1}
& E\left[ {\left( {{\bf{x}} - {\bf{\hat x}}} \right){{\left( {{\bf{x}} - {\bf{\hat x}}} \right)}^T}} \right] = {{\bf{\Sigma }}_{{{\bf{n}}_1}}}\\
\label{err2}
 &= {{\bf{\Sigma }}_{\bf{x|z}}} - {\bf{B}}{{\bf{\Sigma }}_{\bf{u|z}}}{{\bf{B}}^T}\\
\label{err3}
 &= {{\bf{\Sigma }}_{\bf{x|z}}} - {{\bf{\Sigma }}_{\bf{xu|z}}}{{\bf{\Sigma }}_{\bf{u|z}}}^{ - 1}{{\bf{\Sigma }}_{\bf{xu|z}}}^T\\
\label{err4}
&= {{\bf{\Sigma }}_{\bf{x|z}}}\! -\! \left( {{{\bf{\Sigma }}_{\bf{x|z}}} \!-\! {{\bf{\Sigma }}_{\bf{x|yz}}}} \right)\!{{\bf{U}}^T}{\left( {{\bf{\Lambda }} \!+\! {{\bf{\Sigma }}_{\bf{\nu}}}} \right)^{ - 1}}{\bf{U}}\!\left( {{{\bf{\Sigma }}_{\bf{x|z}}} \!-\! {{\bf{\Sigma }}_{\bf{x|yz}}}} \right),
\end{align}

\noindent
where (\ref{err1}) and (\ref{err2}) follow from (\ref{xhat}) and (\ref{estxzu}), (\ref{err3}) is result of substituting (\ref{weird1}) in (\ref{err2}), and (\ref{err4}) follows from (\ref{weird2}) and (\ref{weird5}). Using (\ref{V1}), (\ref{sig1}), (\ref{lambda}) and (\ref{cov_noise}), we can rewrite (\ref{err4}) as follows:

\begin{align}
& E\left[ {\left( {{\bf{x}} - {\bf{\hat x}}} \right){{\left( {{\bf{x}} - {\bf{\hat x}}} \right)}^T}} \right] \nonumber \\
&= {{\bf{\Sigma }}_{\bf{x|z}}} - {{\bf{V}}^{-1}}{\rm{diag}}\left\{\! {{{\lambda _i} \!-\! \min \left({\lambda _i},{\lambda' _i} \right)} } , i=1,...,n_x \! \right\}\!{{{\bf{V}}^{-T}}} \\
\label{Dsatisfied}
&= {{\bf{\Sigma }}_{\bf{x|yz}}} + {{\bf{V}}^{-1}}{\rm{diag}}\left\{\! {{\min \left({\lambda _i},{\lambda' _i} \right)} } , i=1,...,n_x \! \right\}\!{{{\bf{V}}^{-T}}}.
\end{align}

\noindent
From (\ref{sig2}) and (\ref{V2}) we have:

\begin{equation}
\label{D}
{\bf{D}} = {{\bf{\Sigma }}_{\bf{x|yz}}} + {{\bf{V}}^{-1}}{\rm{diag}}\left\{ { {\lambda' _i} } , i=1,...,n_x \! \right\}\!{{{\bf{V}}^{-T}}}.
\end{equation}

\noindent
Comparing (\ref{D}) and (\ref{Dsatisfied}), it is clear that $E\left[ {\left( {{\bf{x}} - {\bf{\hat x}}} \right){{\left( {{\bf{x}} - {\bf{\hat x}}} \right)}^T}} \right] \preceq {\bf{D}}$.

\end{IEEEproof}


\subsection{Lower Bound}
\label{lower}
Let us denote the quantized encoded sequence by $\bf{w}$, and define $\bf{s}$ as ${\bf{s}} \buildrel \Delta \over = {\bf{Cy + Gz}}$. Then from (\ref{estxyz}) we have ${\bf{n}}_2 = {\bf{x}} - {\bf{s}}$. The reconstruction error can be written as:

\begin{equation}
\label{err}
{\bf{x}} - {\bf{\hat x}} = {{\bf{n}}_2} + {\bf{\upsilon }},
\end{equation}

\noindent
where ${\bf{n}}_2$ is the error resulting from irrelevant information in $\bf{y}$
and $\bf{z}$ (remote source coding), and $\bf{\upsilon}$ is the error due to rate constraints. From (\ref{err}) and the fact that ${{\bf{\Sigma }}_{{{\bf{n}}_2}}} = {{\bf{\Sigma }}_{\bf{x|yz}}}$ we have:

\begin{eqnarray}
\label{cov}
{\mathop{\rm cov}} \left( {{\bf{s}} - {\bf{\hat x}}} \right) = {{\bf{\Sigma }}_{\bf{\upsilon }}} 
= E\left[ {\left( {{\bf{x}} - {\bf{\hat x}}} \right){{\left( {{\bf{x}} - {\bf{\hat x}}} \right)}^T}} \right] - {{\bf{\Sigma }}_{\bf{x|yz}}}.
\end{eqnarray}

Starting from (\ref{cov}) we can write the following chain of inequalities:
 
\begin{eqnarray}
\label{bound}
&&\left| {E\left[ {\left( {{\bf{x}} - {\bf{\hat x}}} \right){{\left( {{\bf{x}} - {\bf{\hat x}}} \right)}^T}} \right] - {{\bf{\Sigma }}_{\bf{x|yz}}}} \right| = \left| {{\mathop{\rm cov}} \left( {{\bf{s}} - {\bf{\hat x}}} \right)} \right|  \\
\label{worst_gauss}
&& \ge \frac{1}{{{{\left( {2\pi e} \right)}^{{n_x}}}}}\exp \left[ {2h\left( {{\bf{s}} - {\bf{\hat x}}} \right)} \right] \\
\label{bound1}
&& \ge \frac{1}{{{{\left( {2\pi e} \right)}^{{n_x}}}}}\exp \left[ {2h\left( {{\bf{s}} - {\bf{\hat x}}|{\bf{\hat x}}} \right)} \right]\\
\label{bound2}
&& \ge \frac{1}{{{{\left( {2\pi e} \right)}^{{n_x}}}}}\exp \left[ {2h\left( {{\bf{s}}|{\bf{\hat x}},{\bf{z}}} \right)} \right]\\
\label{bound3}
&& \ge \frac{1}{{{{\left( {2\pi e} \right)}^{{n_x}}}}}\exp \left[ {2h\left( {{\bf{s}}|{\bf{z}}} \right) - 2I\left( {{\bf{s}};{\bf{w}}|{\bf{z}}} \right)} \right]\\
&& = \left| {{{\bf{\Sigma }}_{\bf{x|z}}} - {{\bf{\Sigma }}_{\bf{x|yz}}}} \right|\exp \left[ { - 2I\left( {{\bf{s}};{\bf{w}}|{\bf{z}}} \right)} \right] \nonumber \\
\label{bound4}
&& \ge \left| {{{\bf{\Sigma }}_{\bf{x|z}}} - {{\bf{\Sigma }}_{\bf{x|yz}}}} \right|\exp \left[ { - 2H\left( {\bf{w}} \right)} \right] \\
\label{bound5}
&& \ge \left| {{{\bf{\Sigma }}_{\bf{x|z}}} - {{\bf{\Sigma }}_{\bf{x|yz}}}} \right|\exp \left( { - 2R} \right),
\end{eqnarray}

\noindent
where (\ref{worst_gauss}) is because Gaussian distribution maximizes the differential entropy, (\ref{bound1}) is because conditioning reduces the entropy, (\ref{bound2}) is result of the fact that $h\left( {{\bf{s}} - {\bf{\hat x}}|{\bf{\hat x}}} \right) = h\left( {{\bf{s}}|{\bf{\hat x}}} \right)$ and conditioning reduces the entropy, (\ref{bound3}) can be obtained from the following chain:

\begin{eqnarray}
h\left( {{\bf{s}}|{\bf{\hat x}},{\bf{z}}} \right) &&= h\left( {\bf{s}} \right) - I\left( {{\bf{s}};{\bf{\hat x}},{\bf{z}}} \right) \nonumber \\
&& \ge h\left( {\bf{s}} \right) - I\left( {{\bf{s}};{\bf{\hat x}},{\bf{z}},{\bf{w}}} \right) \nonumber \\
&& = h\left( {\bf{s}} \right) - I\left( {{\bf{s}};{\bf{z}},{\bf{w}}} \right) \nonumber \\
&& = h\left( {\bf{s}} \right) - \left\{ {I\left( {{\bf{s}};{\bf{z}}} \right) + I\left( {{\bf{s}};{\bf{w}}|{\bf{z}}} \right)} \right\} \nonumber \\
&& = h\left( {{\bf{s}}|{\bf{z}}} \right) - I\left( {{\bf{s}};{\bf{w}}|{\bf{z}}} \right),
\end{eqnarray}

\noindent
and (\ref{bound4}) follows from the following inequalities:

\begin{eqnarray}
I\left( {{\bf{s}};{\bf{w}}|{\bf{z}}} \right) &&= H({\bf{w}}|{\bf{z}}) - H({\bf{w}}|{\bf{s}},{\bf{z}}) \nonumber \\
&&\le H({\bf{w}}|{\bf{z}}) \nonumber \\
&&\le H({\bf{w}}),
\end{eqnarray}

\noindent
and thus $ - I\left( {{\bf{s}};{\bf{w}}|{\bf{z}}} \right) \ge  - H({\bf{w}})$. From (\ref{bound5}) we have:

\begin{align}
\label{trouble}
R & \ge \frac{1}{2} \log \left( \frac{\left| {{{\bf{\Sigma }}_{\bf{x|z}}} - {{\bf{\Sigma }}_{\bf{x|yz}}}} \right|}{\left| {E\left[ {\left( {{\bf{x}} - {\bf{\hat x}}} \right){{\left( {{\bf{x}} - {\bf{\hat x}}} \right)}^T}} \right] - {{\bf{\Sigma }}_{\bf{x|yz}}}} \right|} \right) \\
\label{diagonalize}
& = \frac{1}{2} \log \left( \frac{\left| {\bf{\Lambda}} \right|}{\left| {\bf{V}} \left( {E\left[ {\left( {{\bf{x}} - {\bf{\hat x}}} \right){{\left( {{\bf{x}} - {\bf{\hat x}}} \right)}^T}} \right] - {{\bf{\Sigma }}_{\bf{x|yz}}}} \right) {{\bf{V}}^T}\right|} \right)
\end{align}

\noindent
where (\ref{diagonalize}) follows from (\ref{V1}) and (\ref{sig1}). Let us denote the eigenvalues of ${ {\bf{V}} \left( {E\left[ {\left( {{\bf{x}} - {\bf{\hat x}}} \right){{\left( {{\bf{x}} - {\bf{\hat x}}} \right)}^T}} \right] - {{\bf{\Sigma }}_{\bf{x|yz}}}} \right) {{\bf{V}}^T}}$ sorted in order of magnitude by $\mu_{(1)},...,\mu_{(n_x)}$. From the fact that ${ {E\left[ {\left( {{\bf{x}} - {\bf{\hat x}}} \right){{\left( {{\bf{x}} - {\bf{\hat x}}} \right)}^T}} \right]}} \preceq {{\bf{\Sigma }}_{\bf{x|z}}}$ \footnote{Note that ${{\bf{\Sigma }}_{\bf{x|z}}}$ is the covariance of the reconstruction error for zero rate, therefore the reconstruction error cannot be larger than that. However, it does not mean that the distortion constraint has to be restricted to ${\bf{D}} \preceq {{\bf{\Sigma }}_{\bf{x|z}}}$.} and Theorem \ref{th2} we have $\mu_{(i)} \leq \lambda_{(i)}$, $i=1,...,n_x$. From the distortion constraint we have ${\bf{V}} \left( {E\left[ {\left( {{\bf{x}} - {\bf{\hat x}}} \right){{\left( {{\bf{x}} - {\bf{\hat x}}} \right)}^T}} \right] - {{\bf{\Sigma }}_{\bf{x|yz}}}} \right) {{\bf{V}}^T} \preceq {\bf{V}} \left( {\bf{D} - {{\bf{\Sigma }}_{\bf{x|yz}}}} \right) {{\bf{V}}^T}$, which when combined with (\ref{V2}), (\ref{sig2}) and Theorem \ref{th2} yields $\mu_{(i)} \leq \lambda'_{(i)}$, $i=1,...,n_x$. Therefore we can write:

\begin{eqnarray}
\label{mu}
\mu_{(i)} \leq \text{min} \left\{ \lambda'_{(i)}, \lambda_{(i)} \right\}, \, i=1,...,n_x.
\end{eqnarray}

\noindent
From (\ref{diagonalize}) we have:

\begin{eqnarray}
R && \ge \frac{1}{2} \log \left( \frac{ \prod_{i=1}^{n_x} \lambda_{(i)}}{\prod_{i=1}^{n_x} \mu_{(i)}} \right) \nonumber \\
\label{minim}
&& \ge \frac{1}{2}{\sum_{i=1}^{n_x} {\log \left( {\frac{{{\lambda _{(i)}}}}{\text{min} \left\{ \lambda'_{(i)}, \lambda_{(i)} \right\}}} \right)} } \\
\label{end_low_bound}
&& = \frac{1}{2}{\sum_{i=1}^{n_x} {\log \left( {\frac{{{\lambda _{(i)}}}}{{{{\lambda '}_{(i)}}}}} \right)} ^ + },
\end{eqnarray}

\noindent
where (\ref{minim}) follows from (\ref{mu}). The lower bound is established by (\ref{end_low_bound}).


\section{Special Cases}
In this section, we create a link between (\ref{RD}) and (\ref{scalar}), (\ref{trace}), (\ref{rate-rate}) and (\ref{rd-dcc}). We start from the following corollary:

\begin{corollary} \label{corol1}
For ${{\bf{\Sigma }}_{{\bf{x}}|{{\bf{yz}}}}} \prec {\bf{D}} \preceq {{\bf{\Sigma }}_{{\bf{x}}|{{\bf{z}}}}}$ the RDF (\ref{mut-info})--(\ref{constraint}) is given by:

\begin{equation}
\label{rd-dcc2}
{R}\left( {{{\bf{D}}}} \right) = \frac{1}{2}\log \left( {\frac{{\left| {{{\bf{\Sigma }}_{x|{{\bf{z}}}}} - {{\bf{\Sigma }}_{x|{\bf{y}}{{\bf{z}}}}}} \right|}}{{\left| {{{\bf{D}}} - {{\bf{\Sigma }}_{x|{\bf{y}}{{\bf{z}}}}}} \right|}}} \right).
\end{equation}

\end{corollary} 

\begin{IEEEproof}
We will show that in this special case the lower and upper bounds coincide to (\ref{rd-dcc2}). First note that from the assumption, the distortion constraint and (\ref{trouble}) we have:

\begin{equation}
R \ge \frac{1}{2} \log \left( \frac{\left| {{{\bf{\Sigma }}_{\bf{x|z}}} - {{\bf{\Sigma }}_{\bf{x|yz}}}} \right|}{\left| {{\bf{D}} - {{\bf{\Sigma }}_{\bf{x|yz}}}} \right|} \right) \nonumber
\end{equation}

\noindent
which proves the lower bound. From the assumption we can write (\ref{cov_noise}) as:

\begin{align}
\label{apply_min}
{{\bf{\Sigma }}_{\bf{\nu}}}  &= {{\bf{UV}}^{-1}}{\rm{diag}}\left\{\! { \frac{{\lambda _i} {\lambda' _i}}{{\lambda _i} - {\lambda' _i}} } , i=1,...,n_x \! \right\}\!{{{\bf{V}}^{-T}}{{\bf{U}}^{T}}} \\
& = {{\bf{UV}}^{-1}} { {\bf{\Lambda}} {\left( {\bf{\Lambda}} - {\bf{\Lambda}}' \right)}^{-1}  {\bf{\Lambda}}'  } {{{\bf{V}}^{-T}}{{\bf{U}}^{T}}} \nonumber \\
\label{cov_noise2}
& = {\bf{U}}{\left({{{\bf{\Sigma }}_{x|{{\bf{z}}}}} \!-\! {{\bf{\Sigma }}_{x|{{\bf{yz}}}}}}\right)}{{\left( {{\bf{\Sigma }}_{x|{{\bf{z}}}}} \! - \! {\bf{D}} \right)}^{-1}}{\left({{{\bf{D}}} \!-\! {{\bf{\Sigma }}_{x|{{\bf{yz}}}}}}\right)}{\bf{U}^\textit{T}},
\end{align}

\noindent
where (\ref{apply_min}) follows from the fact that ${\lambda' _i} = \min \left({\lambda _i},{\lambda' _i} \right)$, and (\ref{cov_noise2}) follows from (\ref{V1})--(\ref{sig2}). Substituting (\ref{cov_noise2}) in (\ref{sch2}), one can show that $I({\bf{y;u|z}})$ is equal to (\ref{rd-dcc2}). Following the same lines of argument as in Section III-A, one can show that the reconstruction error at the decoder is exactly the same as the target distortion. This completes the proof.

\end{IEEEproof}

Note that Corollary \ref{corol1} is a generalization of (\ref{rd-dcc}). Also note that (\ref{scalar}) immediately follows from (\ref{rd-dcc2}) by setting $n_x=n_y=n_z=1$.

Let us define a subset ${\mathcal{D}}_1$ of the set of covariance distortion constraints $\mathcal{D}$ as all the covariance distortion constraints $\bf{D}$ for which ${\bf{\Sigma}}_1$ and ${\bf{\Sigma}}_2$ in (\ref{sig1}) and (\ref{sig2}) commute. Similarly, we define the subset ${\mathcal{D}}_2$ of $\mathcal{D}$ as the set of all covariance distortion constraints $\bf{D}$ which commute with ${{\bf{\Sigma }}_{\bf{x|yz}}}$. In the sequel, we will provide two propositions which relate our results to (\ref{trace}) and (\ref{rate-rate}). The proofs are left out due to space limitations.






\begin{proposition} \label{prop1}
Minimization of (\ref{rd-dcc2}) over the set of all covariance matrix distortion constraints ${{\bf{\Sigma }}_{{\bf{x}}|{{\bf{yz}}}}} \prec {\bf{D}} \preceq {{\bf{\Sigma }}_{{\bf{x}}|{{\bf{z}}}}}$ in ${\mathcal{D}}_1$ which satisfy the mean-squared error constraint (\ref{mse}) yields (\ref{trace}) and (\ref{water}).
\end{proposition}

\begin{proposition} \label{prop2}
If ${\bf{y}} = {\bf{x}} + {\bf{n}}_1$ and ${\bf{z}} = {\bf{x}} + {\bf{n}}_2$ where the noise terms ${\bf{n}}_1$ and ${\bf{n}}_2$ are white, mutually independent and independent of ${\bf{x}}$, then minimization of (\ref{rd-dcc2}) over the set of all covariance matrix distortion constraints ${{\bf{\Sigma }}_{{\bf{x}}|{{\bf{yz}}}}} \prec {\bf{D}} \preceq {{\bf{\Sigma }}_{{\bf{x}}|{{\bf{z}}}}}$ in ${\mathcal{D}}_2$ which satisfy $\vert {\bf{D}} \vert \leq e^{-2R_I} \vert {{{\bf{\Sigma }}_{\bf{x|z}}}} \vert $ yields (\ref{rate-rate}) and (\ref{mut-inf-water}).
\end{proposition}




\section{Conclusions}
\label{conclusion}
We upper- and lower-bounded the rate-distortion function for the vector Gaussian remote source coding problem with side information at the decoder and covariance matrix distortion constraints. We further studied some special cases where the exact rate-distortion function can be derived. We showed that several results from existing works can be derived and generalized using these special cases. Future work includes  the derivation of the exact rate-distortion function in the general case and also application of the results to the problem of source coding in wireless acoustic sensor networks.



\begin{thebibliography}{1}




\bibitem{Tian}
C. Tian and J. Chen, \emph{Remote vector Gaussian source coding with decoder side information under mutual information and distortion constraints}, \relax IEEE Transactions on Information Theory, vol. 55, no. 10, pp.4676-4680, Oct. 2009.

\bibitem{Jan}
J. \O stergaard and M. S. Derpich, \emph{Sequential Remote Source Coding in Wireless Acoustic Sensor Networks}, \relax European Signal Processing Conference (Eusipco), pp. 1269-1273, Aug. 2012. 

\bibitem{Wornell}
S. C. Draper and G. W. Wornell, \emph{Side information aware coding strategies for estimation under communication constraints}, \relax IEEE Journal on Selected Areas in Communications, vol. 22, no. 6, pp. 1-11, Aug. 2004. 

\bibitem{myDCC}
A. Zahedi, J. \O stergaard, S. H. Jensen, P. Naylor, and S. Bech, \emph{Distributed remote vector Gaussian source coding for wireless acoustic sensor networks}, \relax IEEE Data Compression Conference, Salt Lake City, UT, USA, Mar. 2014. {\tt arXiv:1401.3945 [cs.IT]}

\bibitem{matrix}
D. Bernstein, \emph{Matrix mathematics, theory, facts and formulas}, \relax Princeton University Press, 2nd edition, 2009.

\end{thebibliography}
\end{document}